# An approximation to determine the source of the WOW! Signal

Alberto Caballero[1]

**Abstract:** In this paper it is analysed which of the thousands of stars in the WOW! Signal region could have the highest chance of being the real source of the signal, providing that it came from a star system similar to ours. A total of 66 G and K-type stars are sampled, but only one of them is identified as a potential Sun-like star considering the available information in the Gaia Archive. This candidate source, which is named 2MASS 19281982-2640123, therefore becomes an ideal target to conduct observations in the search for potentially habitable exoplanets. Another two candidate stars have a luminosity error interval that includes the luminosity of the Sun, and 14 candidates more are also identified as potential Sun-like stars, but the estimations on their luminosity were unknown.

**Keywords:** WOW! Signal, SETI, Search for Extraterrestrial Intelligence, interstellar radio message, alien life.

## 1. Introduction

As of October 2020, the WOW! Signal remains the strongest candidate SETI signal. It has been suggested that the signal was produced by hydrogen clouds from Comets 266/P Christensen and P/2008 Y2 (Paris and Davies, 2015). However, this hypothesis has been dismissed by the scientific community, and the source of the signal remains unknown.

Despite the WOW! Signal never repeated, the key aspect was its duration. The signal lasted for 72 seconds, but since this was the maximum amount of time that the Big Ear radio telescope was able to observe, it is likely that the signal would have lasted longer.

The main problem, however, is that the signal never repeated. Follow-up observations of the area conducted by many observatories during several years never detected another signal (Gray and Ellingsen, 2002). Nonetheless, the fact that the signal never repeated, does not necessarily discard that it was produced by extraterrestrial intelligence.

In fact, if we analyse the history of (the few) radio signals that humanity have sent to several targets in the hope of contacting a civilization, none of those transmissions had a long duration or were repeatedly sent for a long time. An extraterrestrial civilization could have opted to behave in a similar manner.

Few attempts have been made to determine the exact location of the WOW! Signal due to the difficulty involved. Despite it was detected in just one of the two feed horns of the radio telescope, the data was processed in a way that does not allow us to determine which of the feed horns actually received the signal.

The other reason that makes difficult to determine the exact source is the high uncertainly in declination: 20 arcminutes. The following image shows an approximation of the two sections of the sky that could contain the source of the signal, each of them with thousands of stars.

[1] alberto.caballero@uvigo.es

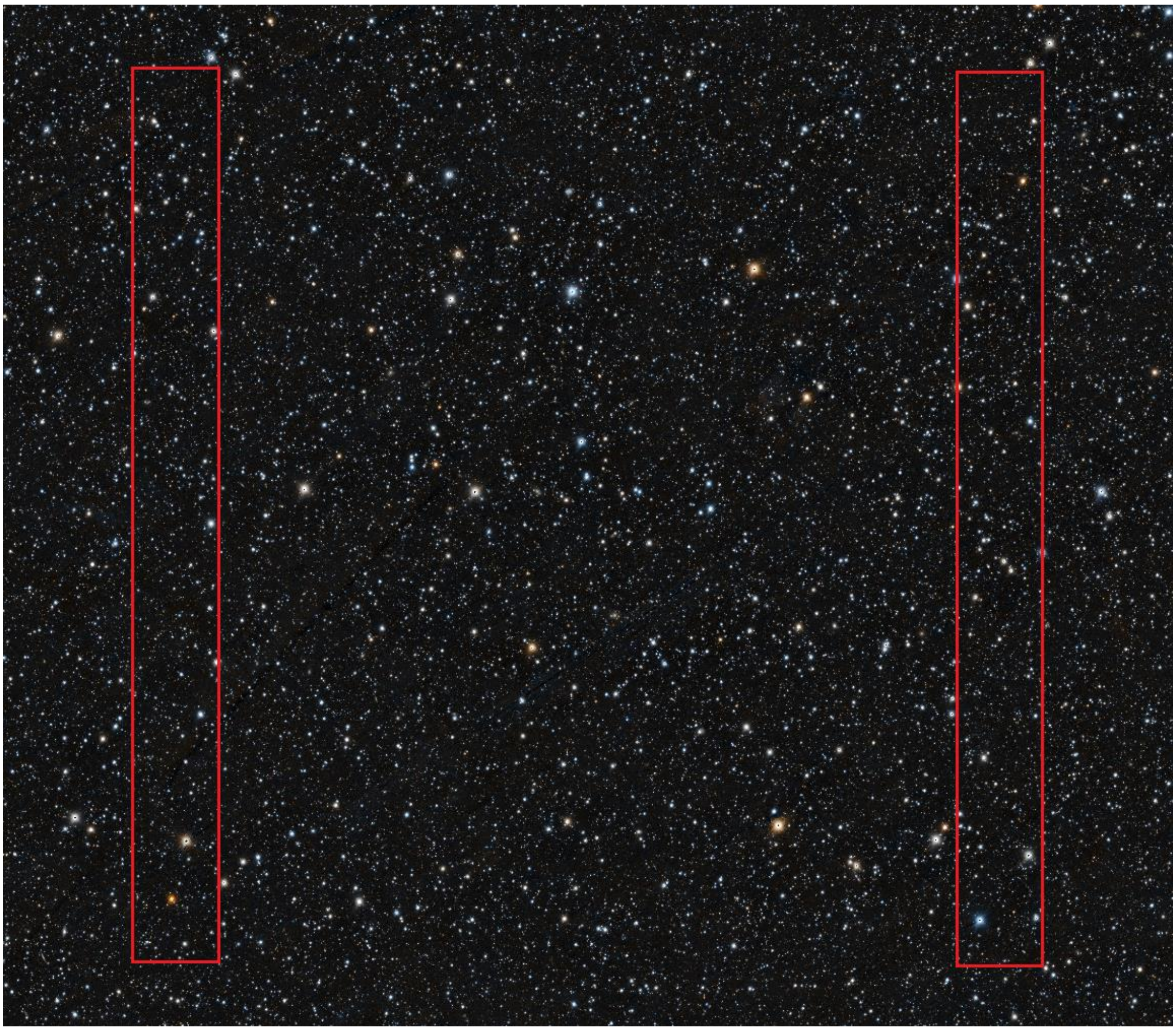

Figure 1: In red, the two regions where the WOW! Signal could have originated
Source: Pan-STARRS/DR1

The coordinates of the signal are RA: 19h25m31s ± 10s (for the positive horn), 19h28m22s ± 10s (for the negative horn), and DEC: −26°57′ ± 20′, both in J2000 equinox (Ehman, 1997). In this article an attempt is made to create a list of the possible sources of the signal assuming that, if it was produced by an extraterrestrial civilization, their exoplanet is similar to Earth.

## 2. Methodology

In order to create a list of possible sources, the Gaia Archive and its 'gaiadr2.gaia_source' database are used. For the positive horn, the RA interval was set between 19:25:21 and 19:25:41. For the negative horn, it was set between 19:28:12 and 19:28:32. The DEC interval was set between -27.2833 and -26.6167 for both horns.

To filter an optimistic sample of stars that range from intermediate K to G type, several parameters were added. Radius_val (estimated radius of the star) was set between 0.83 and 1.15, teff_val (estimated temperature) was set between 4,450 and 6000 Kelvin, and lum_val (estimated luminosity) was set between 0.34 and 1.5.

A conservative sample of candidate sources only include Sun-like stars with a teff_val between 5730 and 5830.

## 3. Optimistic sample of candidate sources

For the positive horn, the following list extracted from the Gaia Archive shows a total of 38 candidates that were found.

| source_id | ra<br>deg | dec<br>deg | parallax<br>mas | phot_g_mean_mag<br>mag | teff_val<br>K | radius_val<br>solRad | lum_val<br>solLum |
|---|---|---|---|---|---|---|---|
| 6765764507112448256 | 291.4078244106725 | -27.25103348064761 | 0.6617190537184244 | 15.817942 | 5338.75 | 1.0674632 | 0.8339861 |
| 6765765258727494400 | 291.3880678944634 | -27.23211134776794 | 0.745458633122556 | 16.440039 | 4985.6665 | 0.8412172 | 0.3939163 |
| 6765764507112457472 | 291.4050378881666 | -27.242314482415665 | 0.8690020759734415 | 15.33458 | 5208.98 | 1.0768784 | 0.7691992 |
| 6765765469185193856 | 291.3405777082392 | -27.231619063047397 | 0.6119436180636963 | 15.834018 | 5791.3335 | 0.95350605 | 0.92141634 |
| 6765765159947515264 | 291.3498168952102 | -27.262044885229212 | 4.132132545543122 | 11.953611 | 5473.5 | 0.9562914 | 0.73949575 |
| 6765776258143127296 | 291.3894525946542 | -27.076231480909186 | 1.039794636307421 | 15.3299265 | 5282.663 | 0.8720368 | 0.53355104 |
| 6765774986832739456 | 291.4063064229606 | -27.143010661715195 | 1.489180692281826 | 14.933055 | 4979.0 | 0.84579146 | 0.39608628 |
| 6765776395582109184 | 291.3549437124267 | -27.071737100248757 | 1.2561866682267304 | 14.481471 | 5354.0 | 1.03361 | 0.79090005 |
| 6765776322562322432 | 291.3802140337682 | -27.067936288917686 | 0.548376333515173 | 15.986092 | 5672.0 | 1.0384039 | 1.0054771 |
| 6765967023410285824 | 291.3471247775688 | -26.937616985335993 | 0.8856357690749613 | 14.837008 | 5739.6665 | 1.0633346 | 1.1055572 |
| 6765967882403793920 | 291.3407598229724 | -26.87675246899194 | 0.834710615463235 | 15.204329 | 5321.6665 | 1.1310799 | 0.9244254 |
| 6765966576733650176 | 291.3815830436253 | -26.948958807736112 | 0.6433463227673782 | 16.145773 | 5100.5 | 1.0541036 | 0.67750466 |
| 6765965850879999232 | 291.36090333085394 | -26.994445033634367 | 0.5845269379581346 | 15.975662 | 5324.0 | 1.13112 | 0.92611355 |
| 6765966542373897984 | 291.4161831365087 | -26.93292040204237 | 2.233219284845393 | 13.188023 | 5360.0 | 1.0520328 | 0.8230239 |
| 6765965820819385088 | 291.3477417977837 | -27.011847032815247 | 0.7358897849096999 | 15.778309 | 5643.3335 | 0.86113155 | 0.677605 |
| 6765966817251822976 | 291.4062167740024 | -26.91715196942292 | 0.8671629077437688 | 15.4946165 | 5597.8447 | 0.8479383 | 0.6360725 |
| 6765966370575194368 | 291.3993763620409 | -26.966779530092996 | 1.9262678052195197 | 14.305581 | 5008.5864 | 0.8598932 | 0.41922235 |
| 6765966576733648768 | 291.37296675485896 | -26.959119803252666 | 0.9686262433653223 | 14.830159 | 5374.5 | 1.131444 | 0.96230567 |
| 6765966503715392256 | 291.4024104176108 | -26.944489145141773 | 2.2952767970537167 | 12.569741 | 5914.5 | 1.0923635 | 1.3155313 |
| 6765970940420526080 | 291.34271484185274 | -26.851809389087308 | 1.015469282933235 | 14.987729 | 5309.6724 | 1.0327581 | 0.76377034 |
| 6765967813684287616 | 291.35558214834754 | -26.8975149203459 | 0.7729907465387106 | 15.615973 | 5348.0 | 0.9988009 | 0.7352214 |
| 6765966576733650816 | 291.3791057796132 | -26.95007259508542 | 0.8129311982244052 | 15.814709 | 4993.0 | 1.0249876 | 0.58827204 |
| 6765995194097657728 | 291.39061240901964 | -26.774968594519052 | 0.7900374481004636 | 16.382814 | 4922.0 | 0.8422995 | 0.37514076 |
| 6765991418824567296 | 291.37685980536037 | -26.852073066964092 | 1.1644933594104796 | 14.982484 | 5093.495 | 0.998458 | 0.6045301 |
| 6765991384464827648 | 291.36886654958215 | -26.861500548892298 | 0.6195416137772571 | 16.585733 | 5202.6665 | 0.851454 | 0.4785431 |
| 6765996400986745856 | 291.3605625401642 | -26.719099268419004 | 0.5490298668754247 | 16.114515 | 5334.6665 | 1.1243402 | 0.92239994 |
| 6765998084613926400 | 291.40933867337293 | -26.669647247919453 | 1.0428271108018765 | 15.674025 | 4867.0 | 0.91072005 | 0.41928554 |
| 6765991040867405568 | 291.3959560903157 | -26.88825362294387 | 0.9275737800195296 | 15.949698 | 4889.0 | 0.891221 | 0.40883276 |
| 6765994579920532224 | 291.37891647634257 | -26.803075768824147 | 0.6165501479055495 | 15.979432 | 5374.5 | 1.0470474 | 0.8240994 |
| 6765996023029593472 | 291.3685770130869 | -26.750913371240294 | 0.7499241878171775 | 15.627783 | 5162.0 | 1.1145408 | 0.7946184 |
| 6765994648640031104 | 291.33770132924815 | -26.818556732487767 | 0.6723169607891949 | 15.748095 | 5332.0 | 1.0882161 | 0.8623538 |
| 6765994339402347136 | 291.3522928418148 | -26.852118591986823 | 0.7348805803778272 | 16.118809 | 4870.0 | 1.051287 | 0.560084 |
| 6765996882023035392 | 291.4096869651997 | -26.731824228151872 | 0.5614153597745786 | 15.911455 | 5452.0 | 1.1477177 | 1.0485479 |
| 6765997633639271552 | 291.399361758971 | -26.732110164267503 | 0.7219073480322802 | 16.138458 | 5339.5 | 0.8439137 | 0.5215466 |
| 6765996126108802816 | 291.38633280833335 | -26.741936856548573 | 1.1596343970316858 | 15.29579 | 5107.935 | 0.86186033 | 0.4555646 |
| 6765990284953127808 | 291.41888355191026 | -26.90546674512369 | 1.0200372317532642 | 15.25953 | 5086.6636 | 1.006677 | 0.6112334 |
| 6765991143946632064 | 291.3971991733755 | -26.86656692552351 | 1.6675611560931494 | 14.08374 | 5312.75 | 0.9523269 | 0.65094507 |
| 6766010316680810880 | 291.40816971816764 | -26.634124358309908 | 0.5060977859031384 | 16.258465 | 5733.0 | 0.96940476 | 0.9146033 |

Figure 2: List of G and early-to-mid K type stars in the WOW! Signal region, positive feed horn
Source: ESA

With the parallax values is possible to know how far each star is. The higher the parallax, the closer the star. Out of the 38 stars, the closest one seems to be source_ID: 6765765159947515264, with a parallax of 4.132132545543122 milliarcseconds, which is around 242 parsecs, or 789 light years. For the negative

horn, the following list extracted from the Gaia Archive shows a total of 28 candidates that were found.

| source_id | ra deg | dec deg | parallax mas | phot_g_mean_mag mag | teff_val K | radius_val solRad | lum_val solLum |
|---|---|---|---|---|---|---|---|
| 6765736057248561280 | 292.1006411084745 | -27.21261451522549 | 0.6770301939417982 | 15.823268 | 5769.5 | 0.87326604 | 0.76127326 |
| 6765743066636007296 | 292.09496370141426 | -26.99746405912848 | 0.851001741594187 | 15.519336 | 5400.25 | 0.9272313 | 0.65875864 |
| 6765741314288690304 | 292.060361979927 | -27.0593914881932 | 1.6625818516315762 | 14.252357 | 5144.3335 | 0.9551288 | 0.5756189 |
| 6765730319172282240 | 292.0560940194295 | -27.224607313260446 | 0.6585505506342175 | 15.246085 | 5827.0195 | 1.1461735 | 1.3645244 |
| 6765742615659064320 | 292.0929014486811 | -27.04707574266142 | 0.5619097663211164 | 16.100163 | 5774.6665 | 0.9244046 | 0.8561041 |
| 6765742856177287296 | 292.0868209180845 | -27.031051726628903 | 1.1606470433537 | 14.985551 | 5230.0 | 0.93771935 | 0.59271675 |
| 6765740111697782016 | 292.09474860429157 | -27.108498876731172 | 0.6230461047113127 | 16.299883 | 5234.28 | 0.95176905 | 0.6126121 |
| 6765735228318190080 | 292.12287647730295 | -27.22957020598409 | 0.6959928922176472 | 16.41032 | 4993.0 | 0.91000867 | 0.4636946 |
| 6765730353532007808 | 292.07902880171923 | -27.21720126238249 | 1.1529966721926206 | 14.469317 | 5424.5 | 1.0984589 | 0.94124174 |
| 6765742894836663296 | 292.0759513968115 | -27.030785816974635 | 0.6098307913751307 | 16.160534 | 5499.3335 | 0.92362803 | 0.70295763 |
| 6765740042978281600 | 292.1265748541325 | -27.10483117369423 | 0.7224034322118007 | 16.033098 | 5072.115 | 1.0025636 | 0.59934235 |
| 6765790345636231552 | 292.05900196055336 | -26.960555365946902 | 5.99738492568554 | 10.847295 | 5438.6665 | 1.1128556 | 0.9762072 |
| 6765790379995741568 | 292.1087982137261 | -26.97226361335549 | 0.6850750555485354 | 15.4897785 | 5871.3335 | 0.9688844 | 1.0050453 |
| 6765790861032086912 | 292.0930345124132 | -26.915360128567457 | 0.6568770612523586 | 16.530819 | 4838.0 | 0.9899087 | 0.48366922 |
| 6765791548226860032 | 292.0526515365629 | -26.92236323264193 | 0.5711494805445938 | 16.207207 | 5220.0 | 1.0906745 | 0.7957325 |
| 6765790581857735424 | 292.1253427521075 | -26.944584694882142 | 0.6072714874748856 | 16.341255 | 5171.467 | 0.98649144 | 0.62709934 |
| 6766173662876487040 | 292.1018977249007 | -26.703533903204548 | 0.8196983473427807 | 15.744305 | 5295.955 | 0.9085708 | 0.5850452 |
| 6766185860583649280 | 292.07907228629284 | -26.65468402891079 | 0.8343066865491464 | 15.039279 | 5727.6934 | 1.0330772 | 1.0348547 |
| 6766170570500493184 | 292.0538562962534 | -26.818120313593298 | 1.16983605540771 | 15.3357725 | 4952.4927 | 0.9067327 | 0.44560352 |
| 6766170364342062720 | 292.0638655786175 | -26.824618062756596 | 0.5808766874796816 | 15.902043 | 5872.9 | 0.9445489 | 0.9562116 |
| 6766174040833637248 | 292.09353773619483 | -26.664311851782127 | 1.3169351714788855 | 13.793874 | 5763.25 | 1.1457651 | 1.3048353 |
| 6766174075193374464 | 292.1042995558813 | -26.655675199554384 | 1.3838029577182769 | 14.053238 | 5342.5 | 1.1486186 | 0.9683326 |
| 6766185791864654720 | 292.082562206216 | -26.670163196643735 | 1.8108125151055896 | 13.3890915 | 5783.0 | 0.9965662 | 1.0007366 |
| 6766167237604777728 | 292.0821866791453 | -26.865065174203693 | 0.6077678961151007 | 16.136177 | 5231.75 | 1.05332 | 0.7488647 |
| 6766167271964515456 | 292.09507515404727 | -26.861305154776026 | 0.946032078198879 | 15.3182125 | 5293.3335 | 0.9590382 | 0.6505543 |
| 6766167271964515712 | 292.0928906903116 | -26.856928874197855 | 1.0494598017362597 | 14.216689 | 6000.0 | 1.085432 | 1.3756406 |
| 6766166962726868864 | 292.10542793511127 | -26.89363989479729 | 0.6353080864011327 | 15.979972 | 5447.0 | 0.9848016 | 0.7691689 |
| 6766166206812624000 | 292.1162400523931 | -26.901633336763922 | 0.8649620494184755 | 15.2424755 | 5457.8867 | 1.0112089 | 0.8174752 |

Figure 3: List of G and early-to-mid K type stars in the WOW! Signal region, negative feed horn
Source: ESA

In this sample, the closest star appears to be source_ID: 6765790345636231552, with a parallax of 5.99738492568554 milliarcseconds, which is around 166 parsecs, or 544 light years. The fact that all the stars in both samples are farther than 500 light years away is consistent with Claudio Maccone estimations that the closest communicative civilization is no closer than 500 light years away (Maccone, 2012).

Maccone points out that the distance in which the existence of a communicative civilization is more likely is 1,933 light years away, which is 592 parsecs, equivalent to 1.68 millarcseconds. In the sample of the positive feed horn, the star (apparently a G-type) closer to that distance is source_id: 6765741314288690304, located 1,961 light years away. This star has an estimated temperature 634 Kelvin lower than the Sun, a radius 5% lower, and a luminosity 43% lower.

In the sample of the negative horn, the star (apparently a K-type) closest to that distance is source_id: 6765991143946632064, located 1,955 light years away. This star has an estimated temperature 466 Kelvin lower than the Sun, a radius 5% lower, and a luminosity 35% lower.

## 4. Conservative sample of candidate sources

If we introduce the interval values corresponding to Sun-like stars in the Gaia Archive, no stars are found in the positive horn beam. If we introduce the same intervals for the negative horn beam, only one Sun-like star is found: source_id: 6766185791864654720, with a RA of 292.082562206216, and a DEC of -26.670163196643735.

The star, which is named as 2MASS 19281982-2640123 in the 2MASS archive, has an estimated temperature of 5,783 Kelvin, a radius of 0.9965662 solar radii, and a luminosity 1.0007366 times that of the Sun. It has a parallax of 1.81 milliarcseconds, which is 552 parsecs, or 1,801 light years.

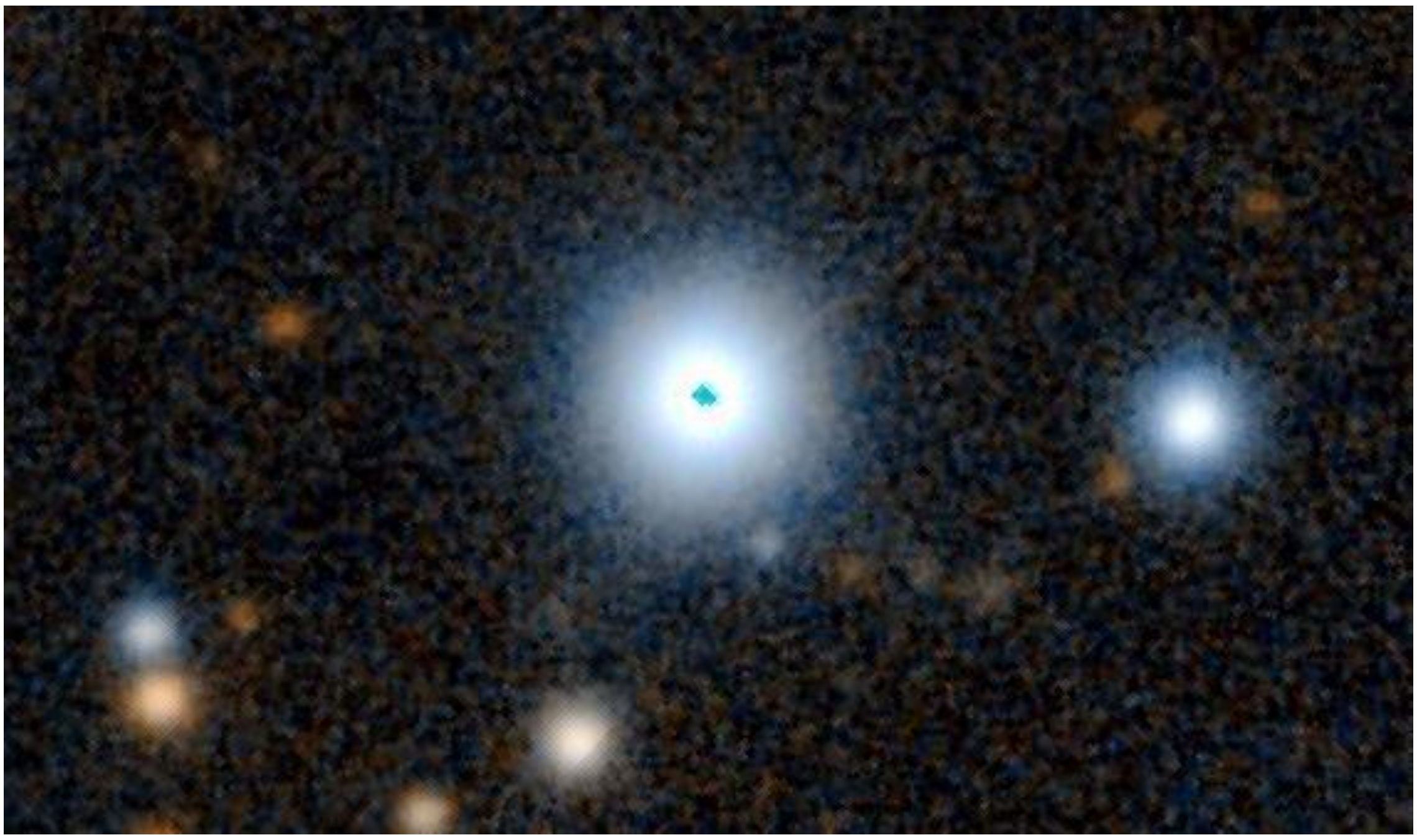

Figure 4: The only potential Sun-like star found in the WOW! Signal region with the available data
Source: PanSTARRS/DR1

2MASS 19281982-2640123 could be, therefore, the only Sun-like star found among the thousands of stars located in the WOW! Signal region. The location of this star has a RA error of 2.2 seconds and a DEC error of 17 arcminutes with respect to the signal.

There are also two more conservative candidates with estimated temperatures similar to the Sun, and with the luminosity of the Sun falling within the interval of their estimated luminosity lower and upper errors. The first one is source_id 6765765469185193856, which corresponds to 2MASS 19252173-2713537, with an estimated temperature of 5,791 K and a luminosity of 0.92 Solar lum. The second one is source_id 6765742615659064320, which is 2MASS 19282229-2702492, with an estimated temperature of 5,774 K and luminosity of 0.85.

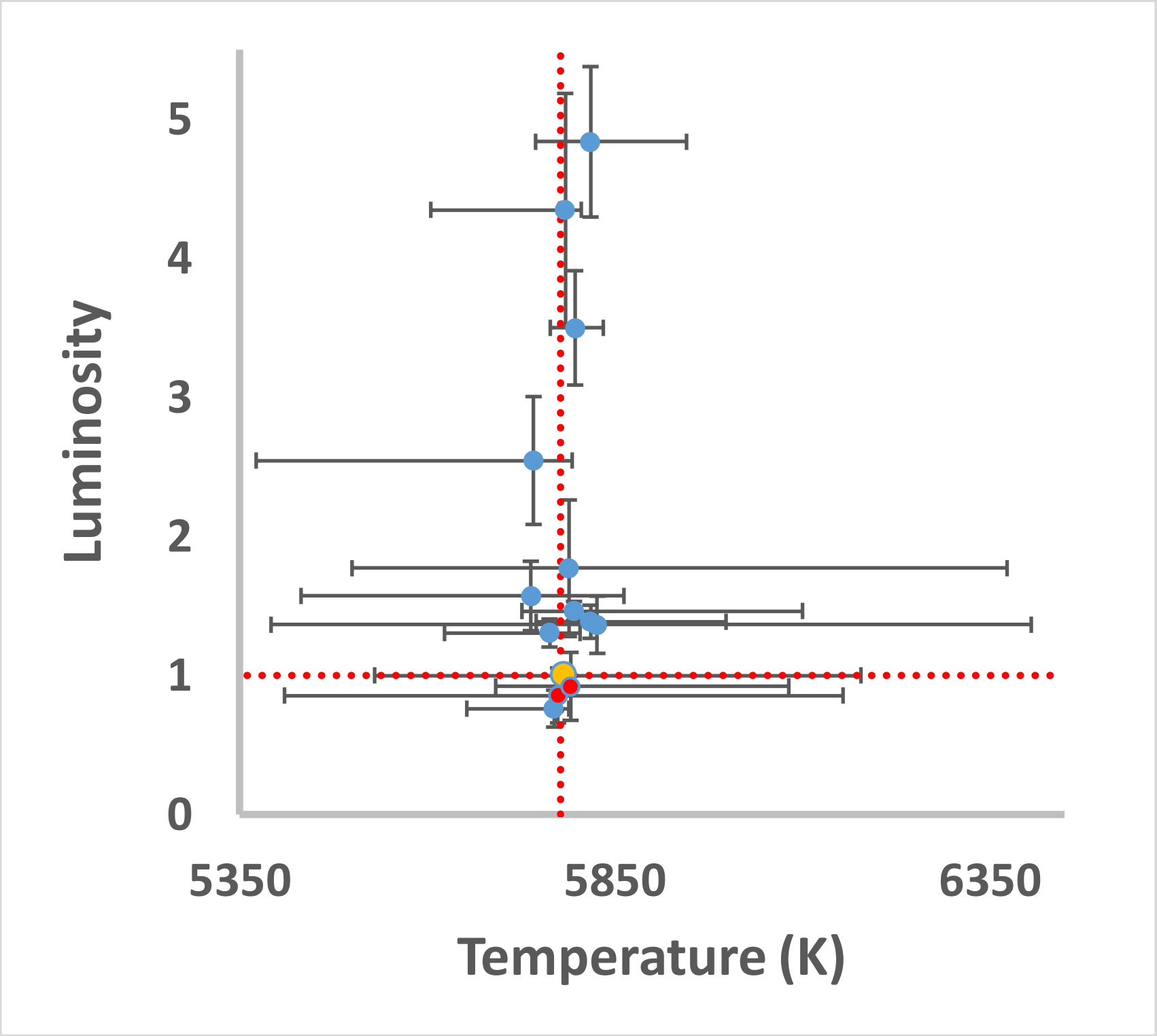

Figure 5: Candidate stars with estimated temperatures between 5,730 and 5,830 K – Error shown for both luminosity and temperature – Vertical red line corresponds to Solar luminosity, and the horizontal red line to Solar temperature – Yellow dot corresponds to 2MASS 19281982-2640123, red dots to 2MASS 19252173-2713537 and 2MASS 19282229-2702492 (below).

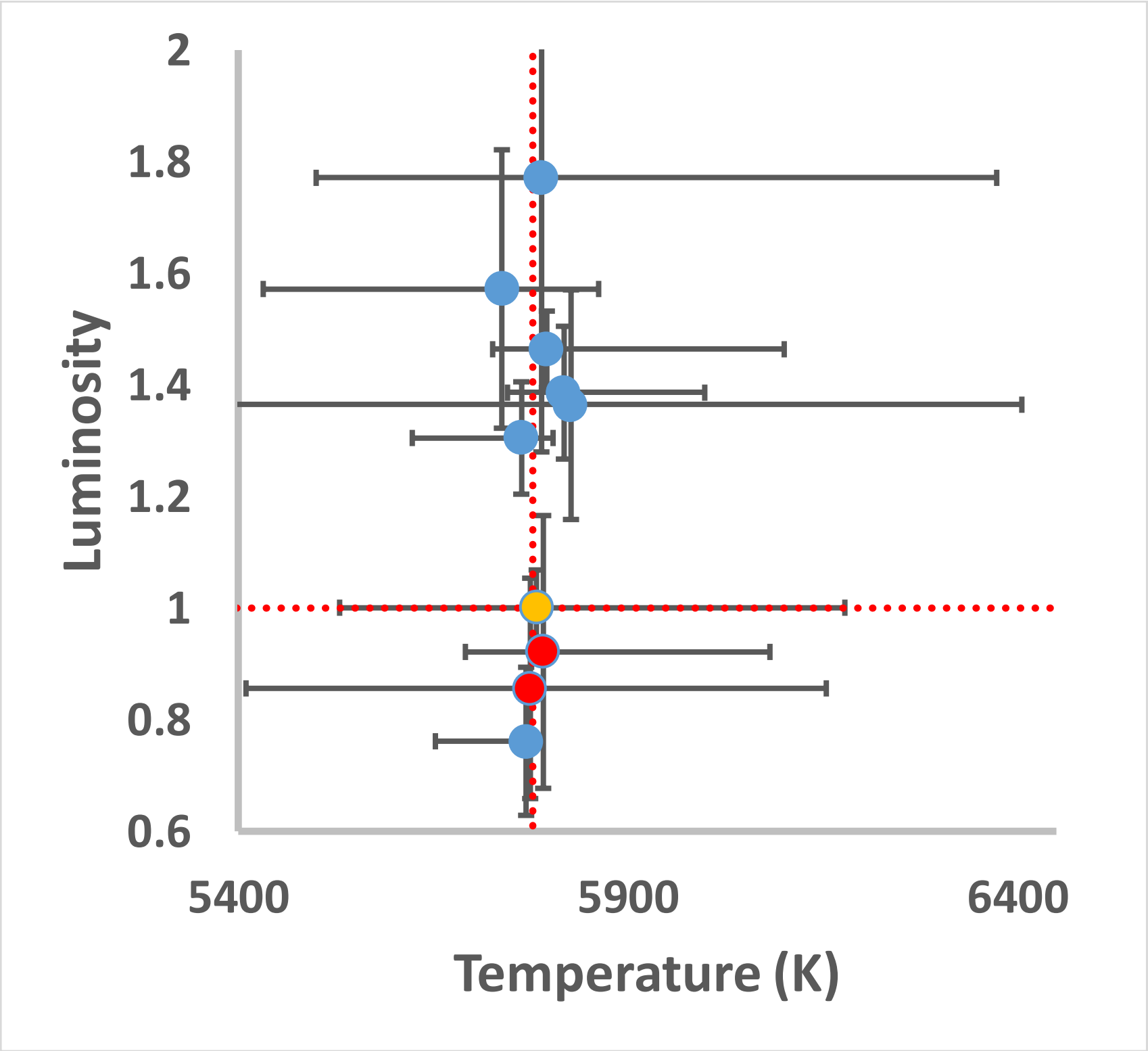

Figure 6: This is figure 5 zoomed in on the candidates with a luminosity more similar to the Sun.

Apart from the three candidates mentioned above, in the Gaia Archive another 14 potential Sun-like stars in the WOW! Signal region were found. However, there is no available data about their luminosity and radius.

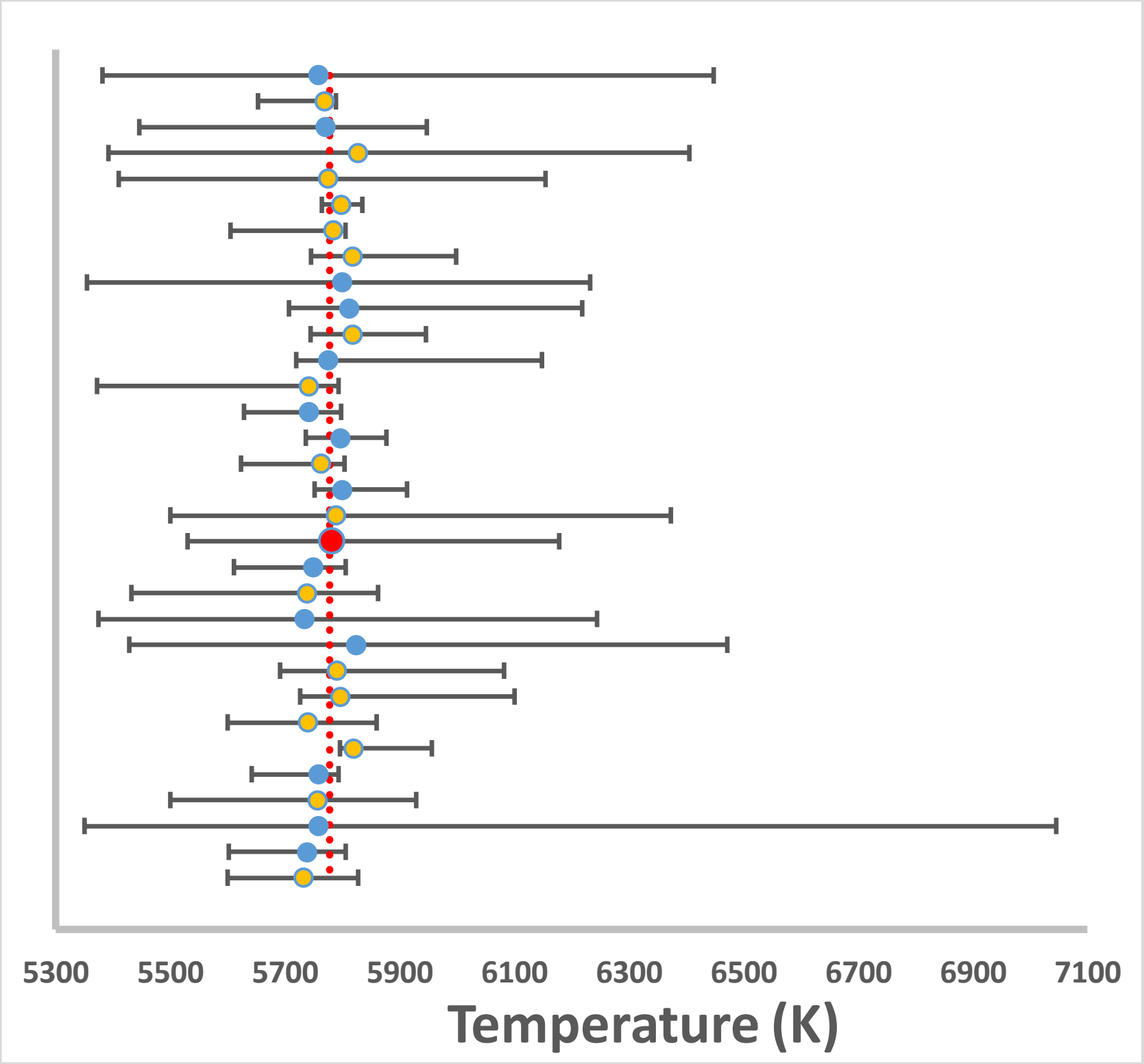

Figure 7: All conservative candidate stars with estimated temperatures between 5,370 and 5,830 K, including errors – Vertical red line corresponds to Solar temperature – Red dot is 2MASS 19281982-2640123, yellow dots are the stars with known temperature and luminosity, and blue dots are the stars with only known temperature.

Some of the stars without known luminosity could be potential Sun-like stars. The most interesting ones are those with an estimated temperature similar to the Sun, and a low error in this parameter. They are source_id 6766167787361134720 (2MASS 19283011-2649582), source_id 6766167168886391296 (2MASS 19282778-2652013), and source_id 6765739355783521280 (2MASS 19281550-2709296).

## 5. Conclusions

In this paper several candidate sources for the WOW! Signal have been suggested. In the region ranging from 19h25m31s ± 10s to −26°57′ ± 20′, and 19h28m22s ± 10s to −26°57′ ± 20′, a total of 66 G and K-type stars were found in the Gaia DR2 archive. Out of this sample, two stars are close to the celestial distance with the highest chance of having a communicative civilization, according to Maccone's mathematical estimations.

With the available data, the only potential Sun-like star in all the WOW! Signal region appears to be 2MASS 19281982-2640123. Despite this star is located too far for sending any reply in the form of a radio or light transmission, it could be a great target to make observations searching for exoplanets around the star.

However, more information such as metallicity, age, and presence or not of stellar companions is needed in order to determine that 2MASS 19281982-2640123 is indeed a Sun-like star. Moreover, another two candidate stars have an error interval of their luminosity that covers the luminosity of the Sun, and three candidates more among 14 are also identified as potential Sun-like stars, but the estimations on their luminosity were unknown.

It is also important to mention that the signal could have come from any of the 66 G and K-type stars, a star that only meets one or two of the parameters set for the optimistic sample (in the WOW! Signal region, a total of 550 stars with a temperature between 4,450 and 6,000 K were found, but no information about their luminosity and radius is available), stars that are not included in the Gaia Archive, a star that is too dim to image with current technology, an extragalactic source, or any other origin.

In any case, since all these stars are located in the same part of the sky, it is ideal to search for exoplanets in the whole region where the WOW! Signal could have come from.